\newcommand{\CLASSINPUTtoptextmargin}{0.75in}
\newcommand{\CLASSINPUTbottomtextmargin}{1.01in}
\documentclass[conference]{IEEEtran}
\IEEEoverridecommandlockouts

\usepackage{cite}
\usepackage{multicol}

\usepackage{amssymb}

\usepackage{amsmath}

\usepackage{amsfonts}
\usepackage[utf8]{inputenc}
\usepackage[T1]{fontenc}
\usepackage{algpseudocode}

\usepackage[english]{babel}
\usepackage[usenames, dvipsnames]{color}
\usepackage{pifont}
\usepackage{booktabs}
\usepackage{mathtools}
\usepackage{textcomp}
 \usepackage{graphicx}
\usepackage{subfigure}
\usepackage[ruled,vlined,commentsnumbered]{algorithm2e}
\usepackage{amsthm}
\usepackage{setspace}
\usepackage{fancyhdr}
\fancypagestyle{AcceptanceSentence}{%
  \lhead{This work has been accepted for publication in the IEEE GLOBECOM 2026 Conference.}
}

\SetKwInput{KwIn}{Inputs} 
\DeclarePairedDelimiter{\ceil}{\lceil}{\rceil}

\theoremstyle{definition}

\theoremstyle{remark}

\theoremstyle{plain}

\newcommand{\RNum}[1]{\uppercase\expandafter{\romannumeral #1\relax}}
\newcommand{\forallK}{\forall\,k \in \mathcal{K}}

\def\BibTeX{{\rm B\kern-.05em{\sc i\kern-.025em b}\kern-.08em
    T\kern-.1667em\lower.7ex\hbox{E}\kern-.125em}}

\begin{document}
\title{SARAMS: Split-Aware Joint Resource Allocation\\for Multi-Monostatic Sensing}

\author{
Maximiliano Rivera Figueroa$^{*}$, Pradyumna Kumar Bishoyi$^{\dagger}$, and Marina Petrova$^{*}$\\[2pt]
$^{*}$Chair of Mobile Communications and Computing (MCC), RWTH Aachen University, Aachen, Germany\\
$^{\dagger}$Department of Electrical Engineering, Indian Institute of Technology Jodhpur, Rajasthan, India\\
Emails: \{maximiliano.rivera@mcc., petrova@mcc.\}rwth-aachen.de, pradyumna@iitj.ac.in
}

\maketitle
\thispagestyle{AcceptanceSentence}
\begin{abstract}

Cooperative multi-monostatic sensing enables sub-meter passive localization in integrated sensing and communication~(ISAC) networks by fusing base station~(BS) observations at a central unit~(CU). Existing studies, however, treat the fused data as ideally available at the CU, overlooking how the per-BS 3GPP functional split jointly constrains fronthaul bitrate, computational load, and whether coherent or non-coherent fusion is feasible at the CU. We propose the \textit{Split-Aware Joint Resource Allocation for Multi-Monostatic Sensing}~(SARAMS) algorithm, which minimizes the worst-case multi-target squared position error bound~(SPEB) by jointly optimizing per-BS power, bandwidth, and observation time under fronthaul, computational, and power constraints. A split-dependent coefficient embeds the feasible fusion type into the Fisher information matrix~(FIM), casting SPEB minimization as a mixed-integer nonlinear program (MINLP) solved via a semidefinite program for power allocation and block coordinate descent (BCD) over a dominance-pruned configuration set. Simulation results demonstrate that SARAMS reduces the 90th-percentile worst-case SPEB by $65.7\%$ over equal power allocation while attaining an $8.7\%$ optimality gap relative to exhaustive search.

\end{abstract}

\begin{IEEEkeywords}
ISAC, Passive sensing, Functional Splits, Resource Orchestration, SPEB, Multi-Monostatic Sensing
\end{IEEEkeywords}

\section{Introduction} \label{sec:Introduction}

Integrated sensing and communication~(ISAC) has emerged as a key enabler for 6G perceptive mobile networks, where cellular infrastructure reuses spectrum, hardware, and resources to simultaneously sense the environment and communicate~\cite{Zhang2021-PMN}. Achieving the sub-meter localization accuracy required for safety-critical applications such as autonomous driving and industrial automation~\cite{Rel19.22.837} is challenging with a single base station~(BS), as passive target accuracy is tightly bounded by propagation conditions and available frequency-time sensing resources.

Multi-monostatic sensing addresses these limitations by exploiting geometric diversity across cooperating BSs. Each BS transmits and receives its own monostatic sensing signal, and a fusion center combines the resulting observations~\cite{figueroa2023cooperative}. 
The authors in~\cite{Behdad2024} study multi-static target detection and power allocation for ISAC in cell-free massive multiple-input multiple-output~(MIMO). Joint sensing-communication beamforming under a similar architecture is investigated in~\cite{Demirhan2025}, and power allocation for interference mitigation in multi-monostatic sensing is studied in~\cite{figueroa2023jcs}. However, these works assume ideal data access at the fusion center, ignoring the constraints of the radio access network (RAN). In the next-generation RAN (NG-RAN), baseband processing is distributed across the central unit~(CU), distributed unit~(DU), and radio unit~(RU) via standardized functional split options~\cite{Rel18.38.401}, which determine what data is forwarded over the fronthaul, constraining bitrate, computational load distribution, and the type of data available at the fusion center. 
These architectural considerations are critical for multi-monostatic sensing, as coherence-preserving splits forward complex-valued channel data enabling coherent fusion across BSs~\cite{Wymeersch2022PartI}, whereas higher-layer splits that forward only magnitude-domain outputs restrict the fusion center to non-coherent combining~\cite{Richards2005FundamentalsOR,Ai2015}. Therefore, the functional split creates an architecture-dependent bottleneck that couples fronthaul bitrate, computational load, and sensing accuracy, a coupling that existing cooperative sensing studies do not capture.

Functional split resource orchestration has been studied on the communication side to minimize energy consumption under fronthaul and computational constraints~\cite{Demir2024ORAN}, and networked ISAC works optimize transmission variables under Cram\'er-Rao Bound-based objectives with limited fronthaul~\cite{Zhu2025}. \textit{However, no prior work models the effect of 3GPP functional splits on sensing performance under joint fronthaul and computational constraints.} In particular, the split boundary determines whether coherent or non-coherent fusion is feasible at the CU, a distinction that directly scales each BS's contribution to localization accuracy.

Motivated by this gap, in this paper we develop the \textit{Split-Aware Joint Resource Allocation for Multi-Monostatic Sensing}~(SARAMS) algorithm, which couples the functional split deployed at each BS with the joint allocation of sensing resources. We model how the coherence of the sensing data forwarded under each functional split affects the multi-target squared position error bound (SPEB). Based on this model, we formulate a worst-case multi-target optimization problem that jointly allocates per-BS transmit power, sensing bandwidth, and observation time under fronthaul constraints, RU and DU complexity constraints, per-BS power constraints, and total network power constraints. In particular, we (i) introduce a split-aware SPEB model that captures the impact of heterogeneous functional splits on localization accuracy, quantifying the accuracy penalty at the fusion center when non-coherent fusion is imposed by higher-layer splits, (ii) analyze the computational complexity of the sensing processing chain at the RU and DU for each functional split option, deriving closed-form giga operations per second~(GOPS) expressions as a function of the split, sensing bandwidth, and observation time, and (iii) formulate a min-max multi-target SPEB optimization over per-BS transmit power, sensing bandwidth, and observation time under fronthaul, computational, and power constraints, and solve it via a block coordinate descent (BCD) algorithm that alternates between optimal power allocation for a fixed sensing configuration and a pruned discrete search over bandwidth and observation time configurations.

\section{System Model}\label{sec:SystemModel}

\subsection{Multi-Monostatic Sensing Framework}\label{subsec:MM_Framework}

\begin{figure*}[t]
\centering
\includegraphics[clip, trim=0 0.1cm 0 0.1cm, width=0.95\textwidth]{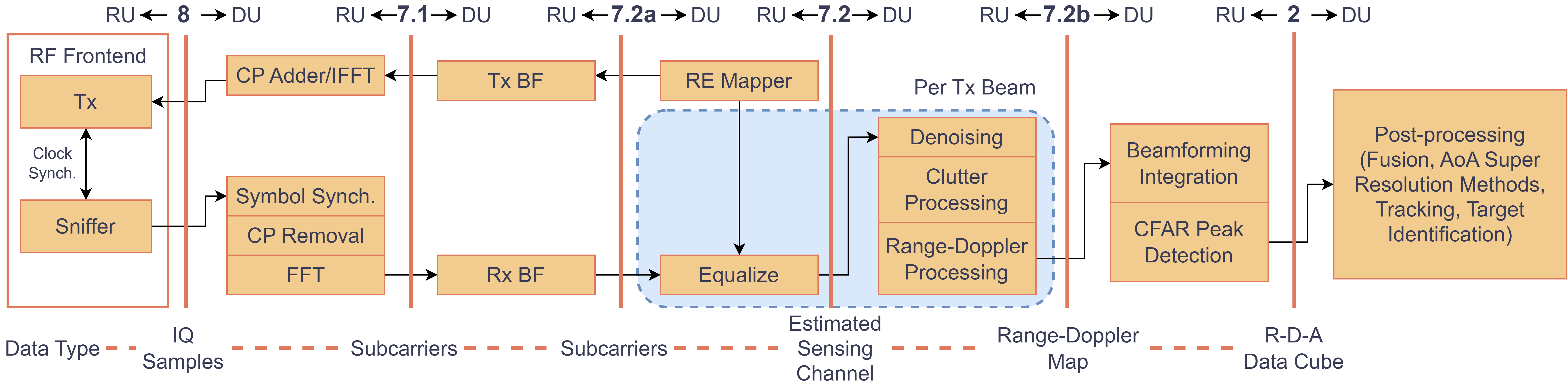}
\caption{Proposed functional splits between RU-DU for sensing services and DU-CU split. CP: cyclic prefix; IFFT: inverse fast Fourier transform; BF: beamforming; RE: resource element; CFAR: constant false alarm rate; R-D-A: Range-Doppler-Angle.}
\label{fig:functional_splits}
\end{figure*}
We consider $K$ BSs deployed in an urban environment, each equipped with a uniform linear array~(ULA) of $N_{\text{tx}}$ transmit and $N_{\text{rx}}$ receive antenna elements operating in full-duplex mode via a co-located sniffer. All BSs have known positions and are time-synchronized and connected to a CU via high-capacity fronthaul links. Each BS transmits an orthogonal frequency division multiplexing~(OFDM) waveform and processes the reflected echo to extract sensing parameters such as time-of-arrival, angle-of-arrival~(AoA), and radial velocity~\cite{figueroa2023cooperative,Wymeersch2022PartI}, which are forwarded to the CU for cooperative localization of $Q$ passive targets at unknown positions $\mathbf{x}_q \in \mathbb{R}^2$, $q \in \mathcal{Q} = \{1, \dots, Q\}$. The form and quality of the forwarded data depend on the functional split at each BS, as discussed in Section~\ref{subsec:Functional_Splits}.
\subsection{Integration in the 5G RAN}\label{subsec:5G_RAN}

The NG-RAN disaggregates gNodeB (gNB) functions into the CU, DU, and RU, and the DU and RU are connected through a common public radio interface (CPRI) or enhanced CPRI fronthaul~\cite{Rel18.38.401}. In multi-monostatic sensing, each BS represents an RU, and the CU serves as the fusion center aggregating observations from all RU-DU pairs.
At each RU, a sniffer captures reflected echo signals and executes the following receiver processing stages, as depicted in Fig.~\ref{fig:functional_splits}: (i)~radio frequency (RF) reception and clock synchronization, (ii)~frame and symbol-level synchronization, (iii)~cyclic prefix~(CP) removal and fast Fourier transform~(FFT) for OFDM demodulation, (iv)~receive beamforming, (v)~zero-forcing~(ZF) equalization to recover the sensing channel, (vi)~denoising and clutter suppression, (vii)~range-Doppler processing via a two-dimensional FFT, (viii)~beamforming integration, and (ix)~constant false alarm rate~(CFAR) peak detection. The point at which this chain is partitioned between RU and DU defines the functional split, which governs both the fronthaul bitrate and the sensing data available at the CU for fusion.

\subsection{Functional Splits}\label{subsec:Functional_Splits}
We adapt the 3GPP functional split options~\cite{Rel18.38.401} to the sensing receiver chain, as illustrated in Fig.~\ref{fig:functional_splits}, where deeper splits assign more processing to the RU at the cost of higher RU computational load. Split~8 forwards raw in-phase and quadrature (IQ) samples at bitrate $R_{\text{S8}}= 2\,N_{\text{ADC}}\,N_{\text{rx}}\,f_s$, where $N_{\text{ADC}}$ is the ADC bit width and $f_s$ the sampling frequency. Split~7.1 adds CP removal and FFT at the RU, giving $R_{\text{S7.1}}= 2\,N_{\text{ADC}}\,N_{\text{rx}}\,N_s/T$, where $N_s$ is the number of active subcarriers and $T$ the OFDM symbol duration. Split~7.2a further applies combiner-based receive beamforming, yielding $R_{\text{S7.2a}} = 2\,\eta_{\text{float}}\,N_s/T$, where $\eta_{\text{float}}$ is the floating-point bit width. Split~7.2 additionally performs ZF equalization with the same bitrate ($R_{\text{S7.2}} = R_{\text{S7.2a}}$). Finally, split~7.2b forwards reduced range-Doppler maps~(RDMs) at $R_{\text{S7.2b}} = \eta_{\text{float}}\,N_{\text{red}}\,M_{\text{red}}\,n_{x,\text{ZP}}\,n_{y,\text{ZP}}/(T\,M\,N_{\text{beam}})$, where $N_{\text{red}}$, $M_{\text{red}}$ are the RDM dimensions, $n_{x,\text{ZP}}$, $n_{y,\text{ZP}}$ zero-padding factors, $M$ the number of OFDM symbols, and $N_{\text{beam}}$ the number of transmit beams. The fronthaul bitrate at BS~$k$ is denoted $R_k$. 

The functional split also determines the quality of sensing data at the fusion center. Splits~8, 7.1, 7.2a, and 7.2 preserve the complex-valued sensing channel, enabling coherent fusion~\cite{Wymeersch2022PartI}, whereas split~7.2b forwards only magnitude information, restricting the CU to non-coherent fusion~\cite{Richards2005FundamentalsOR}. Under heterogeneous deployments, the CU receives a mixture of both data types, reducing the achievable localization accuracy and motivating the split-aware model in Section~\ref{subsec:SPEB}.

\subsection{Computational Load in GOPS}

We quantify the computational load of each processing block in Fig.~\ref{fig:functional_splits} in GOPS following~\cite{Demir2024ORAN}, as summarized in Table~\ref{tab:GOPS}. $N_{\text{FFT}}$ denotes the FFT size, $C_{\text{RF}}$ is modeled as digital baseband filtering with 10 taps~\cite{Demir2024ORAN}, $C_{\text{sync}}$ follows the Schmidl--Cox algorithm~\cite{Schmidl1997}, and $C_{\text{FFT}}$ assumes a split-radix implementation. The beamforming stage follows a combiner-based approach. The range-Doppler periodogram is computed via a two-dimensional FFT over the zero-padded dimensions $N_s' \ge N_s$ and $M' \ge M$~\cite{figueroa2023cooperative}, and cell-averaging CFAR (CA-CFAR) operates over a reduced region defined by $N_{\text{range}} \leq N_s'$ and $M_{\text{speed}} \leq M'$, with $N_c$ reference cells and $N_g$ guard cells. Denoising and clutter suppression are excluded from the GOPS model, since their algorithm choice and complexity vary widely across sensing applications.

\begin{table}[t]
    \centering
    \caption{GOPS per processing block.}
    \label{tab:GOPS}
    \begin{tabular}{ll}
        \toprule
        Block&GOPS \\
        \midrule
        RF processing & $C_{\text{RF}} = 40\, N_{\text{rx}}\, f_s$ \\
        Synchronization & $C_{\text{sync}} = (6\, N_s + 29\, N_{\text{FFT}})/T$ \\
        FFT & $C_{\text{FFT}} = 4\, N_{\text{rx}}\, N_{\text{FFT}} \log_2(N_{\text{FFT}})/T$ \\
        Beamforming & $C_{\text{BF}} = 7\, N_{\text{rx}}/T$ \\
        ZF equalization & $C_{\text{ZF}} = N_s\, N_{\text{beam}}/T$ \\
        Range-Doppler & $C_{\text{RD}} = \frac{(4 N_s' M' \log_2(N_s' M') + 3 N_s' M')\, N_{\text{beam}}}{M\, T}$ \\[1pt]
        CA-CFAR & $C_{\text{CA-CFAR}} =\frac{(N_c - N_g)\, N_{\text{range}}\, M_{\text{speed}}\, N_{\text{beam}}}{M\, T}$ \\
        \bottomrule
    \end{tabular}
\end{table}

Let $C_{\text{RU}}^k$ and $C_{\text{DU}}^k$ denote the aggregate GOPS at the RU and DU when split $f_k$ is deployed at BS~$k$, and let $C_{\text{tot}} = C_{\text{sync}} + C_{\text{FFT}} + C_{\text{BF}} + C_{\text{ZF}} + C_{\text{RD}} + C_{\text{CA-CFAR}}$ be the total sensing processing cost. The RU and DU loads are 
\begin{align}\label{eq:CRU}
    C_{\text{RU}}^k(f_k, b_k, M_k) = C_{\text{RF}} + \Delta C_{\text{RU}}^k(f_k, b_k, M_k), \nonumber\\
    \quad C_{\text{DU}}^k(f_k, b_k, M_k) = C_{\text{tot}} - \Delta C_{\text{RU}}^k(f_k, b_k, M_k),
\end{align} 
where $\Delta C_{\text{RU}}^k$ is the incremental load shifted to the RU: zero for split~8, $C_{\text{sync}} + C_{\text{FFT}}$ for split~7.1, adding $C_{\text{BF}}$ for split~7.2a, $C_{\text{ZF}}$ for split~7.2, and $C_{\text{RD}}$ for split~7.2b. 
The set of RUs connected to DU~$l$ is $\mathcal{K}_l = \{k \in \mathcal{K} : \text{RU}_k \text{ is connected to DU}_l\}$, so the total GOPS at DU~$l$ is $\tilde{C}_{\text{DU}}^{l} = \sum_{k \in \mathcal{K}_l} C_{\text{DU}}^k(f_k, b_k, M_k)$, where $\mathcal{L} = \{1, \dots, L\}$ denotes the set of DUs, and $b_k$ denotes the bandwidth of the $k$-th BS.

\section{Problem Formulation} \label{sec:ProblemFormulation}

\subsection{Squared Position Error Bound and Split-Aware Model}\label{subsec:SPEB}

We quantify the positioning accuracy of the multi-monostatic sensing network through the SPEB, which provides an estimator-independent lower bound on the mean-square localization error~\cite{Shen2010}. For an unbiased estimate $\hat{\mathbf{x}}$ of the target position, the estimation error covariance is lower-bounded by the inverse of the equivalent Fisher information matrix (FIM) $\mathbf{J}_e(\mathbf{x})$~\cite{Shen2010}. The SPEB is defined as
\begin{equation}\label{eq:SPEB}
\mathcal{P}(\mathbf{x}) \triangleq \operatorname{tr}\!\left\{\mathbf{J}_e^{-1}(\mathbf{x})\right\}.
\end{equation}

For the two-dimensional case, the equivalent FIM decomposes into a sum of per-BS contributions 
\begin{equation}\label{eq:FIM}
\mathbf{J}_e(\mathbf{x}) = \sum_{k=1}^{K} \lambda_k \, \mathbf{J}_r(\phi_k),
\end{equation}
where $\phi_k$ denotes the angle from the $k$-th BS to the target and $\mathbf{J}_r(\phi_k)$ is the ranging direction matrix
\begin{equation}\label{eq:Jrphi}
\mathbf{J}_r(\phi_k) = \begin{bmatrix} \cos^2\phi_k & \cos\phi_k\sin\phi_k \\ \cos\phi_k\sin\phi_k & \sin^2\phi_k \end{bmatrix}.
\end{equation}

The scalar coefficient $\lambda_k$ denotes the ranging information intensity of BS~$k$, $\lambda_k = (8\pi^2 b_k^2)(1-\chi_k)\,\mathrm{SNR}_k / c^2$, where $c$ is the speed of light and $\chi_k \in [0,1]$ is the path-overlap coefficient quantifying the ranging information loss due to unresolved multipath~\cite{Shen2010}. The received signal-to-noise ratio (SNR) is $\mathrm{SNR}_k= P_k \, |h_k|^2 \, N_{s,k} \, M_k / \sigma_{n,k}^2$, where $P_k$ is the transmit power, $h_k$ is the complex sensing channel accounting for the target radar cross section (RCS) and round-trip path loss, $N_{s,k}$ is the number of subcarriers, $M_k$ is the number of OFDM symbols, and $\sigma_{n,k}^2$ is the noise power.

The functional split $f_k$ at each BS determines the sensing data type forwarded to the fusion stage~\cite{Wymeersch2022PartI,Richards2005FundamentalsOR}. We denote by $\mathcal{F}_{\mathrm{c}} = \{8,\, 7.1,\, 7.2\text{a},\, 7.2\}$ the set of coherence-preserving splits and by $\mathcal{F}_{\mathrm{nc}} = \{7.2\text{b}\}$ the set of non-coherent splits, and partition the BSs accordingly into $\mathcal{K}_{\mathrm{c}} \triangleq \{k \in \mathcal{K} : f_k \in \mathcal{F}_{\mathrm{c}}\}$ and $\mathcal{K}_{\mathrm{nc}} \triangleq \mathcal{K} \setminus \mathcal{K}_{\mathrm{c}}$. We define the split-dependent coefficient
\begin{equation}\label{eq:beta}
\beta_k(f_k) = \begin{cases} 1, & f_k \in \mathcal{F}_{\mathrm{c}}, \\ \zeta_{\mathrm{nc}}, & f_k \in \mathcal{F}_{\mathrm{nc}}, \end{cases}
\end{equation}
so that the split-aware FIM is given by
\begin{equation}\label{eq:grouped_FIM_compact}
\mathbf{J}_g(\mathbf{x}) = \sum_{k=1}^{K} \beta_k(f_k)\,\lambda_k \, \mathbf{J}_r(\phi_k).
\end{equation}
The penalty factor $\zeta_{\mathrm{nc}} \in (0,1]$ captures the combining loss when the CU is restricted to non-coherent fusion. Since BSs in $\mathcal{K}_{\mathrm{nc}}$ forward amplitude-only range-Doppler maps, the CU combines the $|\mathcal{K}_{\mathrm{nc}}|$ observations without phase alignment, analogous to post-detection pulse integration in radar~\cite{Richards2005FundamentalsOR}, yielding a sublinear effective gain. We model $\zeta_{\mathrm{nc}}$ as the ratio of non-coherent to coherent integration gain relative to a fully coherent system, $\zeta_{\mathrm{nc}} \triangleq G_{\mathrm{nc}}(|\mathcal{K}_{\mathrm{nc}}|)/|\mathcal{K}_{\mathrm{nc}}|$, where $G_{\mathrm{nc}}(n) = \sqrt{n}/\kappa^{\iota(n)}$ is Albersheim's non-coherent integration gain~\cite{Richards2005FundamentalsOR}, with $\kappa = A + 0.12\,A\,B + 1.7\,B$, $A = \ln(0.62/P_{\mathrm{FA}})$, $B = \ln(P_{\mathrm{D}}/(1-P_{\mathrm{D}}))$, and $\iota(n) = 0.454/\sqrt{n+0.44} - 0.38$, where $P_{\mathrm{FA}}$ and $P_{\mathrm{D}}$ denote the false-alarm probability and the detection probability, respectively.
The resulting split-aware SPEB is $\mathcal{P}_g(\mathbf{x}) \triangleq \operatorname{tr}\!\left\{\mathbf{J}_g^{-1}(\mathbf{x})\right\}$.

We extend the single-target SPEB to $Q$ targets at positions $\mathbf{X} = \{\mathbf{x}_1, \dots, \mathbf{x}_Q\}$. Under the target isolation assumption from~\cite{Ai2015}, any two targets $q \neq q'$ are resolvable in range or Doppler at every BS, i.e., $|d_k^{q} - d_k^{q'}| > \Delta r_k$ or $|\nu_k^{q} - \nu_k^{q'}| > \Delta \nu_k$, $\forallK$, where $d_k^q$ and $\nu_k^q$ denote the range and Doppler shift of target~$q$ at BS~$k$, and $\Delta r_k$, $\Delta \nu_k$ are the corresponding range and Doppler resolutions. Under this condition, the FIM block-diagonalizes over targets and the multi-target SPEB is~\cite{Ai2015, Zhu2025}
\begin{equation}\label{eq:SPEB_multi}
\mathcal{P}_{\text{multi-target}}(\mathbf{X}) = \sum_{q \in  \mathcal{Q} } \mathcal{P}_g(\mathbf{x}_q) = \sum_{q \in  \mathcal{Q}} \operatorname{tr}\!\left\{\mathbf{J}_g^{-1}(\mathbf{x}_q)\right\}.
\end{equation}
In the case that targets do not satisfy the target isolation assumption, they are not resolvable and are therefore merged into a single detection that shares the same resources. As the resource allocation algorithm proposed in Section~\ref{sec:ProposedMethod} operates directly on the set of detections the sensing pipeline reports, its allocation procedure remains unaffected regardless of how targets were resolved beforehand.
In the remainder of the paper, superscript~$q$ denotes a per-BS quantity associated with target~$q$, with $\chi_k^q$, $h_k^q$, and $\phi_k^q$ denoting the path-overlap coefficient, the sensing channel, and the AoA of target~$q$ at BS~$k$.

\subsection{Multi-Target SPEB Minimization Problem}

We minimize the worst-case positioning error across all $Q$ targets by jointly optimizing, for each BS $k \in \mathcal{K}$, the transmit power $P_k$, the sensing bandwidth $b_k$ drawn from the set of admissible 5G NR channel bandwidths $\mathcal{B} = \{10, 20, 40, 50, 60, 80, 90, 100\}$~MHz, and the number of OFDM sensing symbols $M_k \in \mathbb{Z}_+$, subject to a total network-wide sensing power budget $P_{\mathrm{tot}}$ shared across all BSs involved in multi-monostatic sensing. The functional splits are assumed fixed and known. 

The network must provide reliable localization for every target observed, including those at unfavorable geometries. A min-max formulation directly enforces a uniform accuracy floor across all $Q$ targets, ensuring that the resource allocation does not neglect the hardest-to-localize target. In contrast, minimizing the sum of SPEBs may concentrate resources on already well-positioned targets at the expense of the worst-served one \cite{figueroa2023jcs}. We therefore adopt the worst-case objective and, using the epigraph reformulation~\cite{Boyd2004} and defining $\xi$ as a slack variable, state the problem as
\begin{align}\label{eq:P1}
    &\min_{\{P_k,\, M_k,\, b_k\}_{k \in \mathcal{K}},\; \xi}  \; \xi \tag{P1}\\
    \textrm{s.t.} \,\, & \operatorname{tr}\!\left\{\mathbf{J}_g^{-1}(\mathbf{x}_q)\right\} \leq \xi,
        \,\,\, \forall\, q \in \mathcal{Q},
        \tag{P1.a}\label{eq:P1a}\\
    &R_k(f_k, b_k, M_k) \cdot M_k/M_{\max} \leq \Gamma,
        \,\,\, \forallK,
        \tag{P1.b}\label{eq:P1b}\\
    &C_{\text{RU}}^k(f_k, b_k, M_k) \leq \bar{G}^{\text{RU}},
        \,\,\, \forallK,
        \tag{P1.c}\label{eq:P1c}\\
    &\tilde{C}_{\text{DU}}^{l} \leq \bar{G}^{\text{DU}},
        \,\,\, \forall l \in \mathcal{L},
        \tag{P1.d}\label{eq:P1d}\\
    &b_k \in \mathcal{B}\, \,\,\, \forallK,
        \tag{P1.e}\label{eq:P1e}\\
    &b_{\min} \leq b_k \leq b_{\max}^k, \,\,\, \forallK,
        \tag{P1.f}\label{eq:P1f}\\
    &M_{\min} \leq M_k \leq M_{\max}, \,\,\, M_k \in \mathbb{Z}_+, \,\,\, \forallK,
        \tag{P1.g}\label{eq:P1g}\\
    &0 \leq P_k \leq P_{\max},
        \,\,\, \forallK,
        \tag{P1.h}\label{eq:P1h}\\
    &\sum_{k \in \mathcal{K}} P_k \leq P_{\mathrm{tot}}.
        \tag{P1.i}\label{eq:P1i}
\end{align}
Constraint~\eqref{eq:P1a} is the epigraph reformulation of the min-max objective. Constraint~\eqref{eq:P1b} enforces per-BS fronthaul capacity over a sensing frame of $M_{\max}$ OFDM symbols. Constraints~\eqref{eq:P1c}--\eqref{eq:P1d} bound the RU and DU computational budgets. Constraints~\eqref{eq:P1e}--\eqref{eq:P1f} restrict $b_k$ to the admissible 5G~NR bandwidth set~$\mathcal{B}$ within $[b_{\min}, b_{\max}^k]$, where $b_{\min}$ reflects the minimum range resolution and $b_{\max}^k$ is the maximum sensing bandwidth allocable at BS~$k$ given its communication load. Constraint~\eqref{eq:P1g} bounds the observation window, where $M_{\min}$ ensures sufficient Doppler resolution and $M_{\max}$ limits the interval over which the target can be assumed static, defining the set $\mathcal{M} = \{M_{\min}, \dots, M_{\max}\}$. Constraints~\eqref{eq:P1h}--\eqref{eq:P1i} enforce the per-BS and total power budgets. Problem~\ref{eq:P1} is a mixed-integer nonlinear program (MINLP), non-convex in the joint decision space, with a combinatorial search over $(b_k, M_k)$ of cardinality $(|\mathcal{B}|\cdot|\mathcal{M}|)^K$, and no closed-form solution.

\section{Proposed Solution}\label{sec:ProposedMethod}

\begin{figure}[t!]
    \centering
    \subfigure[Sensing scenario.]{\includegraphics[clip, trim=4cm 6.5cm 4cm 7cm,
width=0.8\columnwidth]{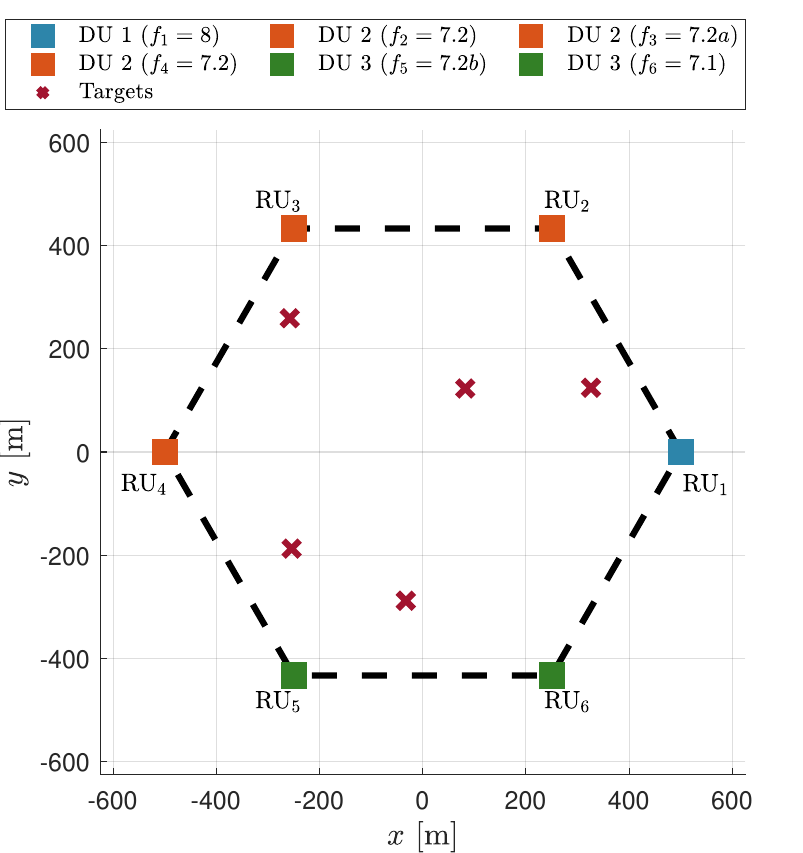}\label{fig:Scenario:a}}

    \subfigure[SPEB vs.\ power budget.]{\includegraphics[clip, trim=1cm 6cm 2cm 7.25cm, width=0.8\columnwidth]{
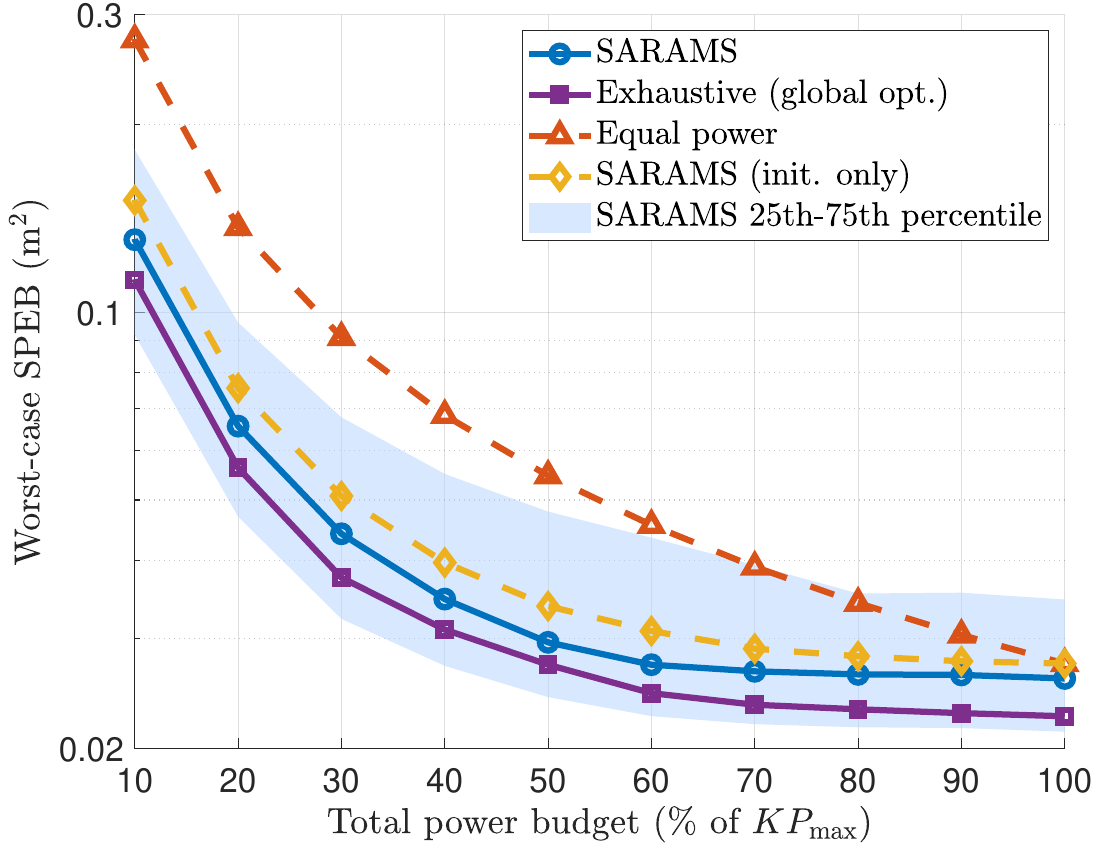}\label{fig:Scenario:b}}

    \subfigure[SPEB vs.\ number of targets.]{\includegraphics[clip, trim=1cm 6cm 2cm 7.25cm, width=0.8\columnwidth]{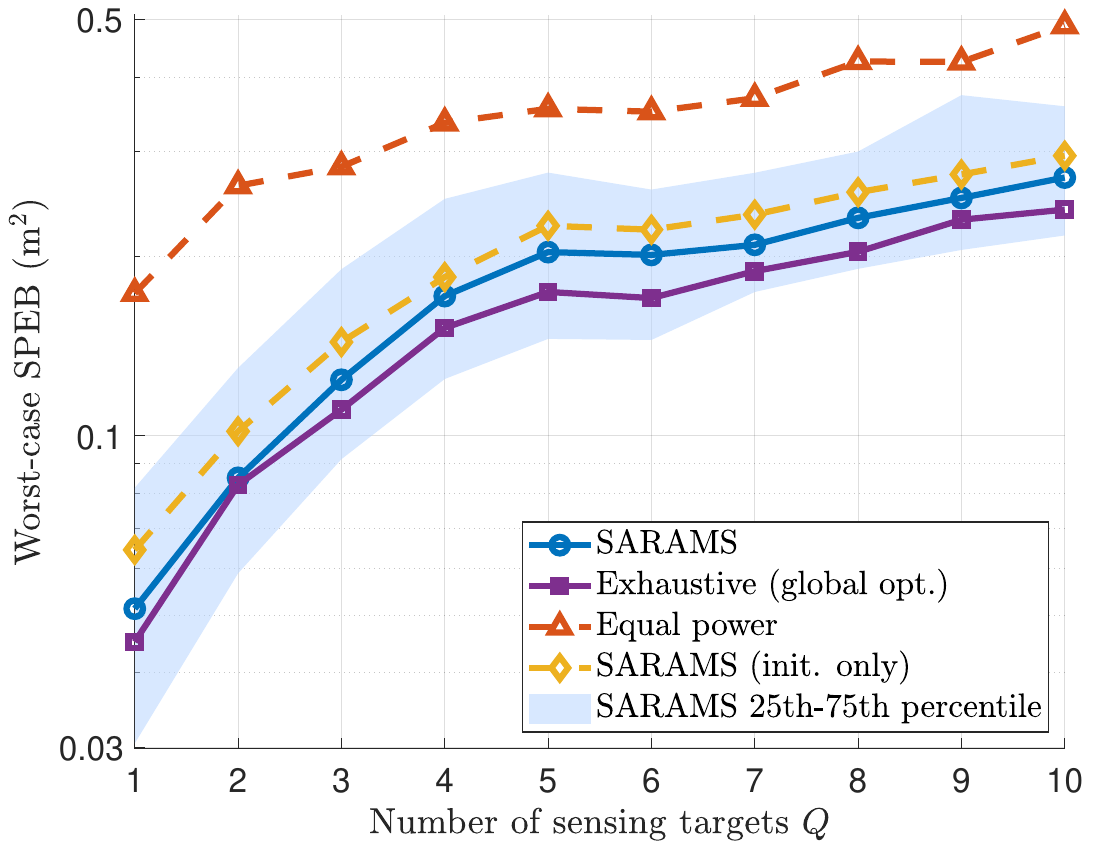}\label{fig:Scenario:c}}

    \caption{(a) Illustration of the sensing scenario, showing $K=6$ RUs (squares) and $Q=5$ randomly placed targets (red crosses). The color code indicates the RU-DU connections, whereas the legend indicates the functional split used. (b) SPEB versus total sensing power budget for $Q = 3$. (c) SPEB versus number of targets for $\alpha = 0.1$.}
    \label{fig:Scenario}
\end{figure}

Problem~\ref{eq:P1} admits a natural decomposition between the continuous power variables
$\{P_k\}_{k \in \mathcal{K}}$ and the discrete configuration variables
$\{b_k,M_k\}_{k \in \mathcal{K}}$. We exploit this structure by defining for
each BS~$k$ the discrete block variable $s_k \triangleq (b_k,M_k)$ and the global
configuration $\mathbf{s} = (s_1,\dots,s_K)$. For any fixed feasible $\mathbf{s}$, we
define
\begin{equation}\label{eq:wkq}
    w_k^q \triangleq \beta_k(f_k)\,\bar{\gamma}_k^q\,M_k\,b_k^3,
    \,\,
    \bar{\gamma}_k^q \triangleq \frac{8\pi^2(1-\chi_k^q)|h_k^q|^2}{c^2\,\Delta f\,\sigma_{n,k}^2},
\end{equation}
so that, for each target $q \in \mathcal{Q}$, the split-aware FIM becomes
\begin{equation}\label{eq:Jg_affine}
    \mathbf{J}_g(\mathbf{x}_q) = \sum_{k=1}^{K} P_k\,\mathbf{A}_k^q,
    \qquad
    \mathbf{A}_k^q \triangleq w_k^q\,\mathbf{J}_r(\phi_k^q) \in \mathbb{S}_+^2,
\end{equation}
where $\mathbb{S}_+^2$ denotes the set of $2 \times 2$ symmetric positive semidefinite matrices. The matrix $\mathbf{J}_g(\mathbf{x}_q)$ is therefore affine in the power vector $\mathbf{p} = [P_1,\dots,P_K]^T$. Since
$\operatorname{tr}(\mathbf{X}^{-1})$ is convex on the positive-definite
cone~\cite{Boyd2004}, \ref{eq:P1} reduces, for fixed $\mathbf{s}$, to a convex power-allocation subproblem.
Applying the Schur complement~\cite{Boyd2004}, we
reformulate it as a semidefinite program (SDP)
\begin{align}
    &\min_{\mathbf{p},\, \xi,\, \{\mathbf{Z}_q\}} \; \xi \tag{P2}\label{eq:P2}\\
    \textrm{s.t.}\quad
    &\begin{bmatrix}
        \sum_{k=1}^{K} P_k\mathbf{A}_k^q & \mathbf{I}_2 \\
        \mathbf{I}_2 & \mathbf{Z}_q
    \end{bmatrix} \succeq 0,
    \quad \forall\, q \in \mathcal{Q}, \tag{P2.a}\label{eq:P2a}\\
    &\operatorname{tr}(\mathbf{Z}_q) \leq \xi,
    \quad \forall\, q \in \mathcal{Q}, \tag{P2.b}\label{eq:P2b}\\
    & \eqref{eq:P1h}-\eqref{eq:P1i}, \nonumber
\end{align}
where $\mathbf{Z}_q \in \mathbb{S}^2$ is an auxiliary symmetric matrix variable for each target~$q$, and $\mathbb{S}^2$ is the set of all $2 \times 2$ symmetric matrices. Each constraint in~\eqref{eq:P2a} is a linear matrix inequality (LMI), since the block matrix depends affinely on the decision variables and is constrained to be positive semidefinite. We denote by $F(\mathbf{s})$ the optimal value of~\ref{eq:P2}, the minimum worst-case SPEB achievable under optimal power allocation for a fixed configuration~$\mathbf{s}$. As~\ref{eq:P2} is a convex SDP, we solve it using CVX.
{\SetCustomAlgoRuledWidth{0.99\columnwidth}
\begin{algorithm}[t]
    \caption{SARAMS: Split-Aware Joint Resource Allocation for Multi-Monostatic Sensing}
    \label{alg:BCD}
    \KwIn{$\{\mathcal{S}_k^{\mathrm{P}}\}_{k \in \mathcal{K}}$, $\bar{G}^{\text{DU}}$, $P_{\max}$, $P_{\mathrm{tot}}$}
    \KwOut{$\mathbf{s}^{\star}$, $\mathbf{p}^{\star}$}

    \For{$k = 1,\dots,K$}{
        $s_k \leftarrow \arg\max_{(b,m)\in\mathcal{S}_k^{\mathrm{P}}} b^3 m$\;
    }
    $\mathbf{s} \leftarrow (s_1,\dots,s_K)$; solve~(P2) to obtain $\mathbf{p}^{\star}$ and $F(\mathbf{s})$\;
    improved $\leftarrow 1$\;

    \While{improved = 1}{
        improved $\leftarrow 0$\;
        \For{$k = 1,\dots,K$}{
            \ForEach{$\tilde{s}_k \in \mathcal{S}_k^{\mathrm{P}}$}{
                $\tilde{\mathbf{s}} \leftarrow (s_1,\dots,s_{k-1},\tilde{s}_k,s_{k+1},\dots,s_K)$\;
                Solve~(P2) for $\tilde{\mathbf{s}}$; obtain $\tilde{\mathbf{p}}$, $F(\tilde{\mathbf{s}})$\;
                \If{$F(\tilde{\mathbf{s}}) < F(\mathbf{s})$}{
                    $s_k \leftarrow \tilde{s}_k$;\, $\mathbf{p}^{\star} \leftarrow \tilde{\mathbf{p}}$;\, $F(\mathbf{s}) \leftarrow F(\tilde{\mathbf{s}})$;\, improved $\leftarrow 1$\;
                }
            }
        }
    }
    \Return{$\mathbf{s}^{\star}=\mathbf{s}$, $\mathbf{p}^{\star}$}\;
\end{algorithm}}

After decoupling the power variables in~\ref{eq:P2}, \ref{eq:P1} is equivalently reduced to a discrete optimization problem
\begin{equation}\label{eq:P3}
    \min_{\mathbf{s} \in \mathcal{S}} \; F(\mathbf{s}), \tag{P3}
\end{equation}
where the per-BS feasible set is
\begin{align}\label{eq:Sk}
    \mathcal{S}_k &\triangleq \bigl\{(b,m) \in \overline{\mathcal{B}}_k \times \mathcal{M} : \nonumber\\
    & R_k(f_k,b,m) \leq \Gamma, \;
    C_{\text{RU}}^k(f_k,b,m) \leq \bar{G}^{\text{RU}} \bigr\},
\end{align}
with $\overline{\mathcal{B}}_k = \{b \in \mathcal{B} : b_{\min} \leq b \leq b_{\max}^k\}$ the set of admissible bandwidths at BS~$k$. Since a DU budget is shared by all RUs connected to it, the DU constraint couples the per-BS configurations, and the joint feasible set is
\begin{equation}\label{eq:Sjoint}
    \mathcal{S} \triangleq \Bigl\{ \mathbf{s} \in \textstyle\prod_{k \in \mathcal{K}} \mathcal{S}_k \; : \; \tilde{C}_{\text{DU}}^{l} \leq \bar{G}^{\text{DU}}, \; \forall l \in \mathcal{L} \Bigr\}.
\end{equation}
We reduce the size of each set $\mathcal{S}_k$ via dominance pruning. Since $\mathbf{J}_g(\mathbf{x}_q)$ depends on $(b_k,M_k)$ only through the product $M_k b_k^3$, given two pairs $(b_1,m_1),(b_2,m_2) \in \mathcal{S}_k$, the latter is dominated whenever $b_2^3m_2 \leq b_1^3m_1$ and $C_{\text{DU}}^k(f_k,b_1,m_1) \leq C_{\text{DU}}^k(f_k,b_2,m_2)$, as it can neither yield a lower SPEB for any power allocation or target nor lower the DU load. Removing all dominated pairs yields the pruned set $\mathcal{S}_k^{\mathrm{P}} \subseteq \mathcal{S}_k$, and $\mathcal{S}^{\mathrm{P}} \triangleq \mathcal{S} \cap \prod_{k \in \mathcal{K}} \mathcal{S}_k^{\mathrm{P}}$ denotes the corresponding pruned joint set.

Despite this reduction, the cardinality of $\prod_{k=1}^{K}\mathcal{S}_k^{\mathrm{P}}$
still grows exponentially with $K$, making exhaustive enumeration intractable.
We therefore propose the SARAMS algorithm, which solves~\ref{eq:P3} via BCD~\cite{Razaviyayn2013BCD},
iteratively updating the discrete allocation of one BS at a time while
holding the remaining allocations fixed and solving~\ref{eq:P2} exactly for each candidate. 
We initialize the algorithm by setting $s_k$ as $\arg\max_{(b,m)\in\mathcal{S}_k^{\mathrm{P}}} b^3m$ for each BS~$k$, reducing the selected configurations along $\mathcal{S}_k^{\mathrm{P}}$ until every DU budget is met, and solving~\ref{eq:P2} to obtain the initial power allocation and $F(\mathbf{s})$.
BSs are then updated cyclically. At each step, every candidate $\tilde{s}_k \in \mathcal{S}_k^{\mathrm{P}}$ is evaluated by solving~\ref{eq:P2} while the remaining configurations are held fixed, and a block update is accepted only upon a strict decrease in $F(\mathbf{s})$. The procedure terminates when a full sweep over all BSs yields no accepted update, as summarized in Algo.~\ref{alg:BCD}.
Since each accepted update strictly decreases $F(\mathbf{s})$ over the finite set $\prod_{k=1}^{K}\mathcal{S}_k^{\mathrm{P}}$, the algorithm is guaranteed to terminate in a finite number of steps. Although global optimality cannot be guaranteed, the final configuration satisfies the coordinatewise optimality condition \cite{Razaviyayn2013BCD}, and no single-BS reconfiguration can further reduce $F(\mathbf{s})$. 

The complexity of Algo.~\ref{alg:BCD} is characterized as follows. In the preprocessing stage, dominance pruning within each $\mathcal{S}_k$ relies on two scalar keys, $b^3m$ and $C_{\text{DU}}^k(f_k,b,m)$. $\mathcal{S}_k^{\mathrm{P}}$ is obtained by sorting the at most $|\mathcal{B}||\mathcal{M}|$ candidates in decreasing order of $b^3m$ and retaining, in a single sweep, those whose DU load falls below the running minimum. Pruning therefore costs $\mathcal{O}(K|\mathcal{B}||\mathcal{M}|\log(|\mathcal{B}||\mathcal{M}|))$ in time and $\mathcal{O}(K|\mathcal{B}||\mathcal{M}|)$ in space, linear in $K$ and independent of $Q$ because $|\mathcal{B}|$ and $|\mathcal{M}|$ are fixed design constants. At each outer sweep, up to $K S_{\max}$ candidate configurations are tested, where $S_{\max}=\max_k |\mathcal{S}_k^{\mathrm{P}}|$, and each test requires solving one SDP. Denoting by $I_{\text{out}}$ the number of outer sweeps until convergence, the overall complexity is $\mathcal{O}\!\left(I_{\text{out}} K S_{\max} T_{\text{SDP}} + K|\mathcal{B}||\mathcal{M}|\log(|\mathcal{B}||\mathcal{M}|)\right)$, where $T_{\text{SDP}}$ denotes the per-SDP cost. Problem~\ref{eq:P2} contains only $K+1+3Q$ scalar variables, $Q$ LMI constraints of size $4 \times 4$, and one linear sum-power constraint, so its per-instance SDP cost remains small for the considered values of $K$ and $Q$. Since pruning runs once while the SDP is solved $I_{\text{out}} K S_{\max}$ times and $T_{\text{SDP}}$ grows with $Q$, the term $I_{\text{out}} K S_{\max} T_{\text{SDP}}$ dominates the overall complexity, and the pruning cost becomes negligible as the number of targets grows.

\section{Performance Evaluation}\label{sec:Results}

\subsection{Setup and Scenario}

To evaluate SARAMS, $K=6$ RUs are deployed at equal angular spacing along a circle of radius $500$~m, each equipped with $N_{\text{tx}}=N_{\text{rx}}=8$ antennas and a $120^\circ$ field of view, forming a hexagonal sensing area. Fig.~\ref{fig:Scenario:a} illustrates the RU-DU distribution and the deployed functional splits. At each Monte Carlo iteration, $Q$ targets are randomly placed within the sensing area. Unless stated otherwise, simulation parameters follow Table~\ref{tab:SimParams}.
SARAMS is benchmarked against three baselines in terms of worst-case SPEB. Exhaustive search over $\mathcal{S}^{\mathrm{P}}$ yields the global optimum of~\ref{eq:P3}. Dominance pruning preserves the minimizer of $F(\mathbf{s})$ over $\mathcal{S}$, and~\ref{eq:P2} is solved exactly for every $\mathbf{s} \in \mathcal{S}^{\mathrm{P}}$. The reported configuration is therefore the global optimum of the joint discrete-continuous problem~\ref{eq:P1}.
Equal power allocation ($P_k = P_{\mathrm{tot}}/K$) quantifies the gain from optimized power distribution. The pre-BCD initialization, where $\mathbf{s}$ maximizes $\sum_{k}b_k^3 M_k$ and~\ref{eq:P2} is solved once without BCD iterations, isolates the contribution of multi-sweep convergence and is referred to as ``SARAMS (init.\ only)'' in the figures.

We conduct two parameter sweeps to assess SARAMS's sensitivity to the power budget and target density. First, the total power budget is varied as $P_{\text{tot}} = \alpha K P_{\max}$, where $\alpha \in (0,1]$ is a power scaling factor. Second, the number of targets $Q$ is varied to assess performance under different target densities. Both sweeps use $1000$ Monte Carlo iterations. We also consider two bandwidth configurations. In the uniform case, all BSs share the same limit $b_{\max}^k = 100$~MHz, $\forallK$. In the heterogeneous case, the per-BS limits are set to $(b_{\max}^1, \dots, b_{\max}^6) = (40, 100, 100, 100, 60, 60)$~MHz, so that constraint~\eqref{eq:P1f} is active for some BSs. This configuration models $\mathrm{RU}_1$, $\mathrm{RU}_5$, and $\mathrm{RU}_6$ as cells subject to higher communication load, which leaves less bandwidth available for sensing.
We assume that the multipath of each target is fully resolvable at every BS, and therefore set $\chi_k^q = 0$, $\forallK$, $\forall q \in \mathcal{Q}$.

\begin{table}[t]
      \centering
      \caption{Simulation Parameters.}
      \label{tab:SimParams}            
      \setlength{\tabcolsep}{3pt}
      \begin{center}
      \begin{tabular}{@{}ll@{\hspace{1.5em}}ll@{}}
          \toprule
          Symbol & Value & Symbol & Value \\
          \midrule
          $f_c$ & 3.5 GHz & $N_{\text{beam}}$ & 17 \\
          $P_{\max}$ & 23 dBm & $N_{\text{red}}$ & 128 \\
          $N_{\text{ADC}}$ & 12 bits & $M_{\text{red}}$ & 128 \\
          $\eta_{\text{float}}$ & 32 bits & $n_{x,\text{ZP}}, n_{y,\text{ZP}}$ & 10 \\
          $N_{\text{FFT}}$ & $2^{\ceil{\log_2(N_s)}}$ & $P_{\mathrm{D}}$ & 0.9 \\
          \text{Noise figure} & 7 dB & $P_{\mathrm{FA}}$ & $10^{-7}$ \\
          $\Delta f$ & 30 kHz & $[M_{\min}, M_{\max}]$ & $[42,\;560]$ \\
          \text{RCS} & 7 dBsm & $\Gamma$ & 500 Mbps \\
          $\bar{G}^{\text{RU}}$ & 90  GOPS &  $\bar{G}^{\text{DU}}$ & 60 GOPS\\
          \bottomrule
      \end{tabular}
      \end{center}
\end{table}

\subsection{Simulation Results and Discussion}

Fig.~\ref{fig:Scenario:b} shows the worst-case SPEB versus the power scaling factor $\alpha$. The SPEB decreases monotonically with $\alpha$, as a higher budget directly increases the per-BS SNR. SARAMS outperforms equal power allocation across all $\alpha$, and the gap with SARAMS (init.\ only) confirms that multi-sweep refinement provides a meaningful gain over the single-pass initialization. SARAMS achieves near-optimal performance relative to the exhaustive search, with a negligible optimality gap across all tested values.

\begin{figure}[t]
    \centering
    \includegraphics[clip, trim=1cm 6.5cm 2cm 7.35cm, width=0.4\textwidth]{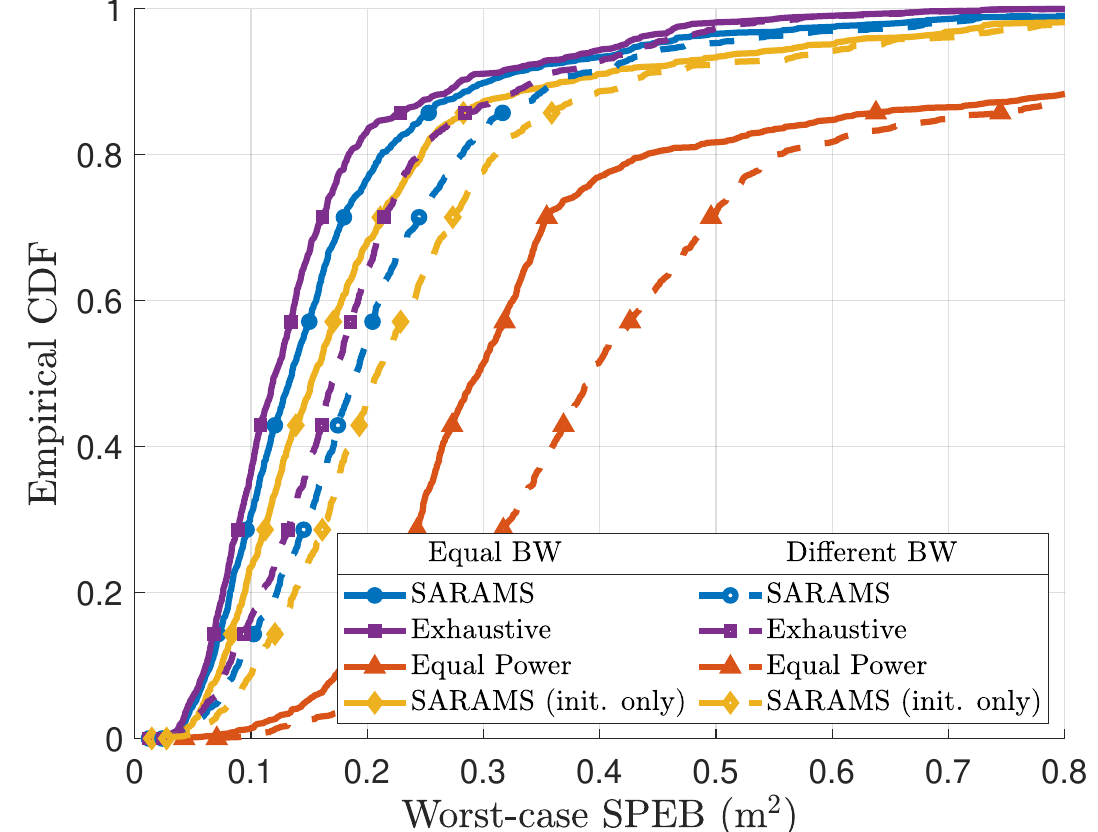}
    \caption{Cumulative distribution function of the worst-case SPEB for $Q=3$ and $\alpha = 0.1$ for different bandwidth configurations.}
    \label{fig:BWComparison}
\end{figure}
\begin{figure}[t]
    \centering
    \includegraphics[clip, trim=1cm 6.6cm 2.5cm 7.23cm, width=0.40\textwidth]{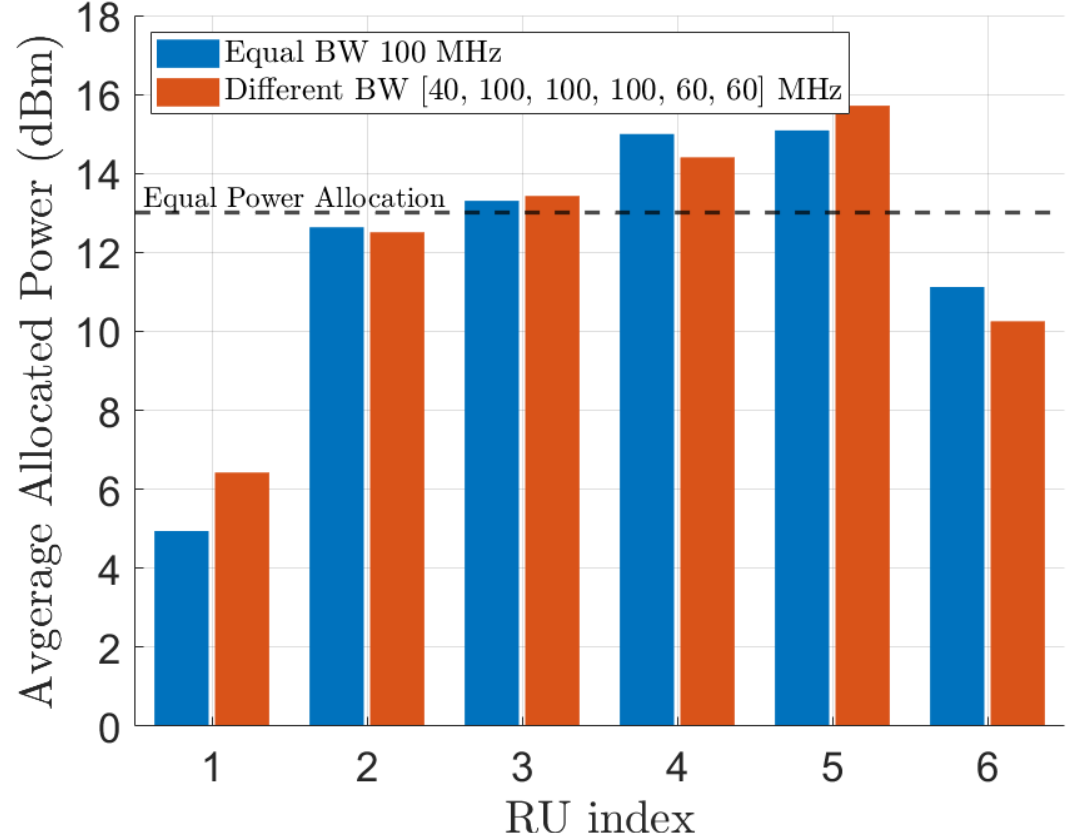}
    \caption{Average power allocation per RU for $Q=3$ and $\alpha = 0.1$ under different bandwidth configurations.}
    \label{fig:PowerAllocationPerRu}
\end{figure}

Fig.~\ref{fig:Scenario:c} shows the worst-case SPEB versus $Q$. As $Q$ increases, the worst-case SPEB grows because a denser target set raises the likelihood of a geometrically unfavorable placement. SARAMS consistently outperforms equal power allocation across all tested values of $Q$, and the gap relative to exhaustive search remains negligible throughout.

Fig.~\ref{fig:BWComparison} shows the cumulative distribution function~(CDF) of the worst-case SPEB for $Q=3$ and $\alpha=0.1$ under both bandwidth configurations. Under the uniform configuration, SARAMS achieves at most $0.3049$~m$^2$ in $90\%$ of cases, outperforming equal power allocation by $65.7\%$ and attaining an $8.7\%$ optimality gap relative to exhaustive search. Under the heterogeneous configuration, tighter per-BS bandwidth limits reduce the available resource space, raising the 90th-percentile SPEB to $0.3632$~m$^2$. Yet SARAMS retains a $60.3\%$ gain over equal power allocation with a $4.8\%$ optimality gap, demonstrating robustness to heterogeneous bandwidth constraints.

Fig.~\ref{fig:PowerAllocationPerRu} depicts the average power allocated by SARAMS per RU under both bandwidth configurations. SARAMS departs from equal power allocation, redirecting the network budget based on the joint heterogeneity of functional split, fronthaul capacity, and target geometry.
$\mathrm{RU}_1$ receives the lowest share in both configurations, since split~8 incurs the highest fronthaul bitrate $R_{\text{S8}}$, binding constraint~\eqref{eq:P1b} and restricting $M_1$. Under the heterogeneous bandwidth setting, the tighter $b_{\max}^1$ relaxes $R_{\text{S8}} \propto f_s \propto b_1$, permitting a larger $M_1$ and a higher power allocation.
In contrast, $\mathrm{RU}_3$ and $\mathrm{RU}_4$ attract a large power share, as their large FIM weight $w_k^q \propto \beta_k\,M_k\,b_k^3$ yields the strongest SPEB contribution.
It is worth noting that a reduced $b_{\max}^k$ has a dual effect on the allocated power, lowering $w_k^q \propto b_k^3$ while relaxing $R_k \propto b_k$, and the balance between these two effects determines whether the allocated power increases or decreases.
\section{Conclusions}\label{sec:Conclusions}
In this paper, we proposed SARAMS, a split-aware resource orchestration algorithm for multi-monostatic sensing under 3GPP functional split constraints. We formulated the joint allocation of per-BS transmit power, sensing bandwidth, and observation time as a min-max multi-target MINLP, incorporating a split-dependent coefficient that scales the per-BS Fisher information contribution according to whether the deployed 3GPP functional split enables coherent or non-coherent fusion at the CU. The problem was decomposed into an SDP for power allocation, solved exactly at each iteration of an outer BCD over a dominance-pruned discrete configuration set. Simulation results demonstrated that SARAMS reduces the 90th-percentile worst-case SPEB by $65.7\%$ over equal power allocation while attaining an $8.7\%$ optimality gap relative to exhaustive search, and that multi-sweep refinement yields meaningful gains over a single-pass initialization.

\section*{Acknowledgment}
This work was funded by the Federal Ministry of Research, Technology and Space Germany within the project ``SENSation'' under grant 16KIS2529.

\bibliographystyle{IEEEtran}
\bibliography{ref.bib}

\end{document}